\documentclass[11pt,showpacs,preprintnumbers,superscriptaddress,amsmath,amssymb,nofootinbib]{revtex4}
\usepackage{graphicx}
\usepackage{dcolumn}
\usepackage{bm}
\usepackage{amssymb}
\usepackage{amsmath}
\usepackage{epsfig}    
\usepackage{color}
\usepackage{slashed}
\usepackage{hhline}
\usepackage{ulem}

\def\be{\begin{equation}}
\def\ee{\end{equation}}
\newcommand{\bea}{\begin{eqnarray}}
\newcommand{\eea}{\end{eqnarray}}

\begin{document}


\title{A radiative lepton model in a non-invertible fusion rule}

\author{Jingqian Chen}
\email{chenjingqian23@mails.ucas.ac.cn}
\affiliation{School of Fundamental Physics and Mathematical Sciences, Hangzhou Institute for Advanced Study, UCAS, Hangzhou 310024, China}
\affiliation{CAS Key Laboratory of Theoretical Physics, Institute of Theoretical Physics, Chinese Academy of Sciences, Beijing, 100190, China}
\affiliation{University of Chinese Academy of Sciences (UCAS), Beijing 100049, China}

\author{Chao-Qiang Geng}
\email{cqgeng@ucas.ac.cn}
\affiliation{School of Fundamental Physics and Mathematical Sciences, Hangzhou Institute for Advanced Study, UCAS, Hangzhou 310024, China}

\author{Hiroshi Okada}
\email{hiroshi3okada@htu.edu.cn}
\affiliation{Department of Physics, Henan Normal University, Xinxiang 453007, China}

\author{Jia-Jun Wu}
\email{wujiajun@itp.ac.cn}
\affiliation{School of Fundamental Physics and Mathematical Sciences, Hangzhou Institute for Advanced Study, UCAS, Hangzhou 310024, China}
\affiliation{University of Chinese Academy of Sciences (UCAS), Beijing 100049, China}
\affiliation{Department of Physics, Henan Normal University, Xinxiang 453007, China}

\date{\today}

\begin{abstract}
{We propose a new mechanism in which electron and muon masses are induced at one-loop level after dynamically violating the symmetry of the Ising fusion rule.
In the neutrino sector, while the neutrino mass is generated at one-loop level, the Ising fusion rule plays a role in stabilizing the particles inside the loop. As a result, the symmetry works like a $Z_2$ symmetry that is not broken at any loop order.
Subsequently, we discuss the lepton flavor violating processes, muon anomalous magnetic dipole moment, and relic density of dark matter, where we specify our dark matter candidate to be a singlet boson and briefly analyze the relic density by estimating the DM annihilation cross section.
Finally, we present results for both the DM cross section and the muon $g-2$, satisfying the neutrino oscillation data as well as the constraints of lepton flavor violations. 
}
 %
 \end{abstract}
\maketitle
\newpage

\section{Introduction}

Recently, attractive symmetries; non-invertible fusion rules which are not groups, are applied to particles physics in which 
they discuss; {\it e.g.},  zero textures to reproduce quark and lepton masses and mixing angles in addition to CP phases in the standard 
model (SM), explanation of the dark matter (DM) candidate~\cite{ Kobayashi:2024cvp,Kobayashi:2025znw, Kobayashi:2025ldi, Choi:2022jqy,Cordova:2022fhg,Cordova:2022ieu,Cordova:2024ypu, Kobayashi:2024yqq, Suzuki:2025oov,Liang:2025dkm, Kobayashi:2025cwx, Heckman:2024obe,Kaidi:2024wio, Kobayashi:2025lar, Nomura:2025yoa}.
But these topics would be  replaced by some flavor symmetries such as (non-)Abelian
discrete (or continuous) groups~\cite{Altarelli:2010gt, Ishimori:2010au, Kobayashi:2022moq, Ishimori:2012zz}.
However, a new unique idea was pointed out by refs.~\cite{Heckman:2024obe, Kaidi:2024wio} in which the symmetry can be dynamically broken at loop-levels even though it is invariant at tree-level. Then, some articles appear, applying these symmetries to phenomenologies~\cite{Suzuki:2025oov, Kobayashi:2025cwx, Nomura:2025sod}.
\footnote{Recently, new types of non-invertible fusion rules were constructed in ref.~\cite{Dong:2025jra} and further intriguing model buildings would be expected.}
Theses symmetries are  supported by some theoretical physics such as Calabi-Yau threefolds \cite{Dong:2025pah}, type-II intersecting/magnetized D-brane models \cite{Kobayashi:2024yqq,Funakoshi:2024uvy}, heterotic string theory on toroidal orbifolds \cite{Dijkgraaf:1987vp,Kobayashi:2004ya, Kobayashi:2006wq, Beye:2014nxa, Thorngren:2021yso, Heckman:2024obe, Kaidi:2024wio},  and world-sheet theory of perturbative string theory \cite{Bhardwaj:2017xup}.

In our paper, we apply the Ising fusion rule (IFR) to the lepton sector in order to induce the masses of electron and muon at one-loop level, while the tauon mass is generated at tree level.
The IFR is given by
\begin{align}
\label{eq:TY}
  \epsilon\otimes\epsilon = \mathbb{I},\quad
  \sigma\otimes\sigma = \mathbb{I}\oplus\epsilon,\quad
    \sigma\otimes\epsilon =  \epsilon\otimes  \sigma=\sigma,
\end{align}
where $\{\epsilon,\sigma \}\otimes \mathbb{I}=\mathbb{I}\otimes \{\epsilon,\sigma\}=\{\epsilon,\sigma\}$.
The masses of electron and muon are forbidden at tree level due to the invertible symmetry, but they are allowed at one-loop level via its dynamically symmetry breaking. 
On the other hand, the neutrino masses are also generated at one-loop level, but its mechanism is the same as the case of an Abelian group such as $Z_2$.
Therefore, the IFR is never broken at any loop order in the neutrino sector.
Also, we discuss the lepton flavor violations (LFVs) of processes such as $\mu \to e\gamma$, muon anomalous magnetic dipole moment (muon $g-2$), and relic density of our DM candidate.
Then, we perform a numerical analysis to satisfy all the data, and briefly show the result.

This paper is organized as follows.
In Sec. \ref{sec:II}, 
we review our lepton seesaw model and explain how to induce small values of the charged-lepton masses as well as the neutrino masses.
Then, we discuss our DM candidate and show the relevant cross sections to explain the relic density.
In Sec. \ref{sec:III}, we demonstrate several numerical results satisfying the neutrino observables, LFVs, muon $g-2$, and relic density of DM in cases of the normal and inverted hierarchies.
In Sec. \ref{sec:IV}, we summarize and conclude.

\section{Model setup}
\label{sec:II}

\begin{table}[t!]
\begin{tabular}{|c||c|c|c|c|c|c||c|c|c|}\hline\hline  
& ~$L_{L_{e,\mu}}$~& ~$L_{L_{\tau}}$~ & ~$\ell_R$~ & ~${E_R}$~ & ~${E_L}$~ & ~${N_R}$~ & ~$H$~ & ~{$\eta$}~ & ~{$S$}~    \\\hline\hline 
$SU(2)_L$   & $\bm{2}$  & $\bm{2}$  & $\bm{1}$ & $\bm{1}$ & $\bm{1}$ & $\bm{1}$  & $\bm{2}$    & $\bm{2}$  & $\bm{1}$  \\\hline 
$U(1)_Y$    & $-\frac12$  & $-\frac12$  & $-1$  & $-1$ & $-1$  & $0$ & $\frac12$  & $\frac12$   & $0$    \\\hline
${\rm IFR}$   & $\bm{\epsilon}$  & $ \bm 1 $& $ \bm 1 $ & $\bm \sigma$ & $\bm \sigma $  & $\bm{\sigma}$ & $\bm{1}$ & $\bm{\sigma}$  & $\bm{\sigma}$          \\\hline 
 \end{tabular}
\caption{Charge assignments of the fields under the $SU(2)_L \times U(1)_Y$ gauge symmetry and the IFR.
}\label{tab:1}
\end{table}

We present our model in this section.
We introduce vector-like fermions $E_{L,R}$ and Majorana right-handed neutral fermions $N_R$, an isospin doublet inert boson $\eta\equiv[\eta^+,(\eta_R+i\eta_I)/\sqrt2]^T$ and a singlet inert boson $S\equiv [S_R + S_I]/\sqrt2$ 
in addition to the SM particles, where all these new particles are assigned to be $\bm \sigma$ under the IFR.

In order to minimally explain the SM lepton masses and mixing angles, we introduce two families of $E_{L,R}$ and $N_R$.
All the SM leptons and Higgs boson are assigned to be trivial singlets under the IFR except the first and second generations of $L_L\equiv [\nu_L,\ell'_L]^T$ that are denoted by $L_{L_{e,\mu}}$.
Here, we impose $L_{L_{e,\mu}}$ to be $\epsilon$ under the IFR.
Due to such a non-trivial assignment for $L_{L_{e,\mu}}$, the terms of $\overline{L_{L_{e,\mu}}} \ell_R H$ are forbidden at tree-level.
Particle contents and their charge assignments are summarized in Table~\ref{tab:1}.
Under these symmetries, we can write the relevant Lagrangian as follows:
\begin{align}
-{\cal L}_\ell &=
\sum_{\ell=e,\mu} \sum_{a=1,2} y_{\ell a} \overline{L_{L_\ell}} \eta E_{R_a} 
+
\sum_{a=1,2} \sum_{i=1}^3  y_{E_{a i}} \overline{E_{L_a}} \ell_{R_i} S
+
 \sum_{a=1}^3  y_{\tau a} \overline{L_{L_\tau}} H \ell_{R_a}
 +
 \sum_{a=1,2}  h_{\tau a} \overline{L_{L_\tau}} \eta E_{R_a}  \label{eq:1-1}\\
&
+
\sum_{\ell=e,\mu} \sum_{a=1,2}y_{\eta_{\ell a}} \overline{L_{L_\ell}} \tilde \eta N_{R_a} 
+
\sum_{a=1,2}y_{\eta_{\tau a}} \overline{L_{L_\tau}} \tilde \eta N_{R_a} 
+
M_{E_a}  \overline{E_{L_a}} E_{R_a}
+
M_{N_a}  \overline{N^C_{R_a}} N_{R_a}
+{\rm h.c.}, \label{eq:1-2}
 \end{align}
where $\tilde\eta\equiv i\tau_2 \eta^*$ with $\tau_2$ being the second Pauli matrix, and
$M_E$ and $M_N$ are diagonal without loss of generality. 
The first two terms in Eq.~(\ref{eq:1-1}) play a role in generating the terms of $\overline{L_{L_{e,\mu}}} \ell_R H$ at one-loop level
with helps of $H^\dag \eta S$ and $M_{E_a}  \overline{E_{L_a}} E_{R_a}$.
Instead of writing down the Higgs potential, we simply extract our essence of this sector by defining the mixing between $S$ and $\eta$ as follows:
\begin{align}
\eta_R &= c_\theta  H_1 +s_\theta H_2,\quad \eta_I = c_\theta  A_1 +s_\theta A_2, \\
S_R &= -s_\theta  H_1 +c_\theta H_2,\quad S_I = -s_\theta  A_1 +c_\theta A_2,
  \end{align}
where we assume the mixing angle for the real part is approximately the same as the one for the imaginary part,
$s_\theta$ is written by $\sim \frac{v_H \mu}{m^2_{H(A)_1} -m^2_{H(A)_2}}$, where $\mu$ is the coefficient of $H^\dag \eta S$
and $m^2_{H(A)_1} ,m^2_{H(A)_2}$ are mass eigenvalues of $H(A)_1,\ H(A)_2$.
In the neutrino sector, the term $\lambda_0(H^\dag \eta)^2$ plays a role in generating the nonzero neutrino mass matrix~\cite{Ma:2006km}.
We will discuss later.

\subsection{Charged-lepton mass matrix}
The charged-lepton mass matrix at tree-level is given by
\begin{align}
{\cal M}_\ell^{\rm tree}=
\frac{v_H}{\sqrt2}
  \left(\begin{array}{ccc} 
0 & 0& 0 \\
0 & 0& 0 \\
y_{\tau1} &y_{\tau2} & y_{\tau3}
  \end{array} \right).
\end{align}
Straightforwardly, we find it we have $m_e=m_\mu=0$, and $m_\tau= \frac{v_H}{\sqrt2} \sqrt{y^2_{\tau1} +y^2_{\tau2} + y^2_{\tau3}}$.
The masses for electron and muon are induced at one-loop level from the following terms:
\begin{align}
&\sum_{\ell=e,\mu} \sum_{a=1,2}  \frac{y_{\ell a}}{\sqrt2} \overline{\ell'_{L_\ell}}  E_{R_a} (c_\theta H_1 +s_\theta H_2) 
+
i \sum_{\ell=e,\mu} \sum_{a=1,2} \frac{y_{\ell a}}{\sqrt2} \overline{\ell'_{L_\ell}}  E_{R_a} (c_\theta A_1 +s_\theta A_2) \\
&+
\sum_{a=1,2} \sum_{i=1}^3  \frac{y_{E_{a i}}}{\sqrt2} \overline{E_{L_a}} \ell_{R_i} (-s_\theta H_1 +c_\theta H_2) 
+
i \sum_{a=1,2} \sum_{i=1}^3  \frac{y_{E_{a i}}}{\sqrt2} \overline{E_{L_a}} \ell_{R_i} (-s_\theta A_1 +c_\theta A_2) 
+{\rm h.c.}, \label{eq:emu_mass},
\end{align}
where we have rewritten them in terms of mass eigenvalues for inert bosons.
Subsequently, the mass matrix at one-loop level is computed by~\footnote{Through the $h$ term in Eq.~(\ref{eq:1-1}), we also obtain the one-loop correction for the $y_{\tau a}$. But we simply neglect this correction.  } 
\begin{align}
(\delta m)_{\ell i}&=
\sum_{a=1,2}  
\frac{y_{\ell a} M_{E_a} y_{E_{a i}} }{2(4\pi)^2}s_\theta c_\theta
\left[
F_I(H_1,H_2,E_a) - 
F_I(A_1,A_2,E_a)
\right], \\
F_I(b_1,b_2,f) &=
\left[
\frac{m^2_{b_1}}{m^2_f - m^2_{b_1}}\ln\left(\frac{m^2_{b_1}}{m^2_f}\right)
-
\frac{m^2_{b_2}}{m^2_f - m^2_{b_2}}\ln\left(\frac{m^2_{b_2}}{m^2_f}\right)
\right],
\end{align}
where $\delta m$ is two by three mass matrix. 
The total charged-lepton mass matrix ${\cal M}_\ell \equiv {\cal M}^{\rm tree}_\ell + \delta m$ is found as
\begin{align}
\frac{v_H}{\sqrt2}
  \left(\begin{array}{ccc} 
\sqrt2 \frac{\delta m_{e1} }{v_H} &\sqrt2 \frac{\delta m_{e2} }{v_H}  &\sqrt2 \frac{\delta m_{e3} }{v_H}  \\
\sqrt2\frac{\delta m_{\mu 1} }{v_H} &\sqrt2\frac{\delta m_{\mu 2} }{v_H}  &\sqrt2\frac{\delta m_{\mu 3 }}{v_H}  \\
y_{\tau1} &y_{\tau2} & y_{\tau3}
  \end{array} \right).
\end{align}
${\cal M}_\ell $ is diagonalized by bi-unitary matrices as $V_L^\dag {\cal M}_\ell V_R =D_\ell= {\rm diag}[m_e,m_\mu,m_\tau]$.
Consequently, we obtain
\begin{align}
{\cal M}_\ell {\cal M}_\ell^\dag
\sim 
 \left(\begin{array}{cc} 
(\delta m \delta m^\dag)_{2\times 2} &(\delta m {\cal M}^{{\rm tree},\dag}_\ell)_{2\times 1}   \\
({\cal M}^{\rm tree}_\ell \delta m^\dag)_{1\times 2} & ({\cal M}^{\rm tree}_\ell {\cal M}^{{\rm tree},\dag}_\ell)_{1\times 1}
  \end{array} \right).
\end{align}
Thus, $m_e,m_\mu \ll m_\tau$ is theoretically explained in diagonalizing the above mass matrix.

\subsection{Active neutrino mass matrix}
The neutrino mass matrix is given at one-loop level~\cite{Ma:2006km}, and the relevant terms are as follows:
\begin{align}
\frac{(y_\eta)_{ia}}{\sqrt2} \overline{\nu_{L_i}} N_{R_a}(c_\theta H_1 + s_\theta H_2)
+
i \frac{(y_\eta)_{ia}}{\sqrt2} \overline{\nu_{L_i}} N_{R_a}(c_\theta A_1 + s_\theta A_2) +{\rm h.c.}
\end{align}
where $i$ runs 1-3, and $a$ runs 1,2.
With help of $(H^\dag \eta)^2$, the neutrino mass matrix is found by
\begin{align}
(m_\nu)_{ij} = \sum_{a=1}^2
\frac{(y_\eta)_{ia} M_{N_a} (y^T_\eta)_{aj}}{2(4\pi)^2}
\left[
c_\theta^2 F_I(H_1,A_1,N) +
s_\theta^2 F_I(H_2,A_2,N)
\right]\equiv  \sum_{a=1}^2
(y_\eta)_{ia} D_{N_a} (y^T_\eta)_{aj} .
\label{eq:neutmass}
\end{align}
Then, $m_\nu$ can be diagonalized by $D_\nu = U^T_\nu m_\nu U_\nu$.
In Eq.~(\ref{eq:neutmass}) and $m_\nu=U^*_\nu D_\nu U^\dag_\nu$, we find the following relation~\cite{Casas:2001sr}
\begin{align}
y_\eta = U^*_\nu \sqrt{D_\nu} O_N \sqrt{D_N^{-1}},
\end{align}
where Max[$|y_\eta|$] has to be less than $4\pi$ which is the perturbative limit.
$O_N$  is 3 by 2 rotation matrix
with an arbitrary complex $z$ satisfying  $O_N^TO_N={1}_{2\times2}$ ($O_NO_N^T={\rm diag}[0,1,1]$ );
\begin{align}
{\rm NH}:\
O_N =\left[\begin{array}{cc}
0 & 0\\
\cos z & -\sin z \\
 \sin z & \cos z \\
\end{array}\right], \quad 
{\rm IH}:\
O_N =\left[\begin{array}{cc}
\cos z & -\sin z \\
 \sin z & \cos z \\
0 & 0\\
\end{array}\right],  
\label{eq:omix}
\end{align}  
where NH(IH) is the short-hand notation of normal (inverted) hierarchy.
The observed mixing matrix $U$ is defined by $V_L^\dag U_\nu$. Therefore, either $V_L$ or $U_\nu$ is known, the other can be expressed in terms of $U$.
The upper bound on sum of neutrino masses; $\sum D_\nu\equiv D_{\nu_1}+D_{\nu_2}+D_{\nu_3}$, is estimated by the minimal cosmological model, that is, $\sum D_{\nu}\le$ 120 meV~\cite{Vagnozzi:2017ovm, Planck:2018vyg}.
Moreover, recent combined data of DESI and CMB suggests us its upper bound be 72 meV~\cite{DESI:2024mwx}.
On the other hand, the lightest mass of our neutrino mass eigenvalues is zero, $\sum D_\nu$ can be written in terms of observables;
\begin{align}
& ({\rm NH}):\   \sum D_{\nu} \sim \sqrt{\Delta m_{\rm atm}^2}\sim{\cal O}(50){\rm meV},\label{eq:mrel}\\
& ({\rm IH}):\   \sum D_{\nu} \sim 2 \sqrt{\Delta m_{\rm atm}^2}\sim{\cal O}(100){\rm meV},
\end{align}
where $\Delta m_{\rm atm}^2\equiv D_{\nu_3}^2 - D_{\nu_1}^2$(NH) and  $\Delta m_{\rm atm}^2\equiv D_{\nu_2}^2 - D_{\nu_3}^2$(IH) are observable and its values are about $2.5\times 10^{-3}$ eV$^2$. 
It implies that both cases satisfy the minimal cosmological model, while only NH satisfies the combined data of DESI and CMB.
%

\subsection{Lepton flavor violations and muon $g-2$}
Here, we discuss the lepton flavor violations (LFVs) and muon $g-2$.
These  dominant forms are derived from the following Lagrangian; 
 \begin{align}
&\frac{Y_{i a}}{\sqrt2} \overline{\ell_{L_i}} E_{R_a}(c_\theta H_1 + s_\theta H_2)
+
i \frac{Y_{i a}}{\sqrt2} \overline{\ell_{L_i}} E_{R_a}(c_\theta A_1 + s_\theta A_2)\\
&
+\frac{G_{a i}}{\sqrt2} \overline{E_{L_a}} \ell_{R_i} (-s_\theta H_1 + c_\theta H_2)
+
i\frac{G_{a i}}{\sqrt2} \overline{E_{L_a}} \ell_{R_i} (-s_\theta A_1 + c_\theta A_2)
+{\rm h.c.},
\end{align}
where $Y\equiv V^\dag_{L} y$, $G\equiv y_E V_R$, and all the above particles are given by mass eigenstates.
Then, the branching ratios for LFVs are found as follows:
  \begin{align}
&{\rm BR}(\ell_i\to \ell_j\gamma)\approx 
\frac{48\pi^3\alpha_{em}C_{ij}}{G^2_F D_{\ell}^2}
\left[
\left| a_{L_{ij}} \right|^2
+
\left| a_{R_{ij}} \right|^2
\right], \\
& a_{L_{ij}}= -\frac{s_\theta c_\theta }{2(4\pi)^2} \sum_{a=1}^2 {Y^\dag_{ai} M_{E_a} G^\dag_{ja}} f(H_{1,2},A_{1,2},E_{a}),\\
& a_{R_{ij}}= -\frac{s_\theta c_\theta }{2(4\pi)^2}\sum_{a=1}^2 {G_{ai} M_{E_a} Y_{ja}}  f(H_{1,2},A_{1,2},E_{a}),\\
& f(H_{1,2},A_{1,2},E_{a})
=F_{LFVs}(H_1,E_a) -F_{LFVs}(H_2,E_a) -F_{LFVs}(A_1,E_a) + F_{LFVs}(A_2,E_a),\\
&
F_{LFVs}(a,b) =
\frac{3 m_a^4 -4m_a^2 m_b^2 +m_b^4+ 2 m_a^4 \ln\left[\frac{m_b^2}{m_a^2}\right]}
{2(m_a^2-m_b^2)^3},
\end{align}
where 
$\alpha_{em}$ is the fine structure constant, $G_F$ is the Fermi constant, and $C_{\mu e}\approx1$, $C_{\tau e}\approx0.1784$, $C_{\tau\mu}\approx0.1736$. 
The most stringent constraint for LFVs is $\mu \to e\gamma$, and the current experimental upper bound of the branching ratio is
$3.1\times10^{-13}$~\cite{MEGII:2023fog}.
And its final sensitivity would reach at $6\times10^{-14}$~\cite{Venturini:2024keu}.
The other upper bounds on the branching ratios are given by~\cite{ParticleDataGroup:2024cfk}
\begin{align}
{\rm BR}(\ell_\tau\to \ell_e\gamma)\lesssim 3.3\times10^{-8},\quad
{\rm BR}(\ell_\tau\to \ell_\mu\gamma)\lesssim 4.2\times10^{-8}.
\end{align}
The recent measurement for the muon anomalous magnetic dipole moment  is close to the prediction of the SM~\cite{Muong-2:2025xyk,Aliberti:2025beg} which is given within 1$\sigma$;
\begin{align}
\Delta a_\mu \simeq (39\pm 64)\times 10^{-11}. \label{eq:mg2_exp}
\end{align}

\subsection{Dark matter candidate}
We now discuss our DM candidate.
In general, we have two DM candidates; the lightest neutral fermion $N_{R_1}$, the lightest inert boson among $H_{1,2},\ A_{1,2}$. 
Here, we concentrate on $H_2(\equiv \chi)$ that is our DM candidate. 
In this case, the dominant cross section to explain the relic density of DM is induced by the following terms:
 \begin{align}
\frac{c_\theta}{\sqrt2} Y_{ia}  \overline{\ell_{L_i}} E_{R_a} \chi
-
\frac{s_\theta}{\sqrt2} G_{ai}  \overline{E_{L_a}} \ell_{R_a} \chi
+{\rm h.c.}.
\end{align}
Then, 
the cross section, which is denoted by $\sigma v_{rel}$, is expanded by the relative velocity $v_{\rm rel}\approx 0.2$ in our galaxy as 
\[\sigma v_{\rm rel}\approx a_{\rm eff} + b_{\rm eff} v_{\rm rel}^2 +{\cal O}(v_{\rm rel}^4), \]
where
\begin{align}
a_{\rm eff}
&\approx  
\frac{(s_\theta c_\theta)^2}{8\pi} \sum_{i,j=1}^3
\left|\sum_{a=1}^2 
\frac{Y_{i,a} M_{E_a} G_{a,j}}{M_{E_a}^2 + m^2_\chi}
\right|^2, \\
b_{\rm eff}
&\approx 
\frac{(s_\theta c_\theta)^2}{24\pi} \sum_{i,j=1}^3
\left|\sum_{a=1}^2 
\frac{Y_{i,a} M_{E_a} G_{a,j}}{(M_{E_a}^2 + m^2_\chi)^2} m_\chi \sqrt{3 M^2_{E_a} + m^2_\chi}
\right|^2 \label{eq.p}.
\end{align}
 The typical range of the cross section to describe the correct relic density is simply estimated as 
 \begin{align}
 1.77552\le (\sigma v_{\rm rel}) \times 10^9\ {\rm GeV}^2 \le 1.96967.\label{eq:relic_exp}
 \end{align}
Here, it corresponds to $0.1196\pm2\times0.0031$~\cite{Planck:2013pxb} at 2$\sigma$.

 \section{Numerical results}
 \label{sec:III}
In this section, we present our numerical results,  applying for the best fit values of the neutrino oscillation data in Nufit 6.0~\cite{Esteban:2024eli}, upper bounds on LFVs, the $\Delta a_\mu$ constraint, and the relic density of DM. In our study, we cut the data of muon $g-2$ if its value is less than $5\times10^{-11}$.
We randomly select our input parameters in the range of $10^2-10^5$ GeV for all the mass parameters,
and $[0-\pi]$ for all the absolute values of our dimensionless parameters.
Note here that once the DM mass is fixed, all masses of the other new particles are greater than $m_\chi$.
Furthermore, $m_{A_1}\sim m_{H_1}$ is assumed simply to evade constraints of oblique parameters.

\subsubsection{{\rm NH}}

\begin{figure}[tb]\begin{center}
\includegraphics[width=88mm]{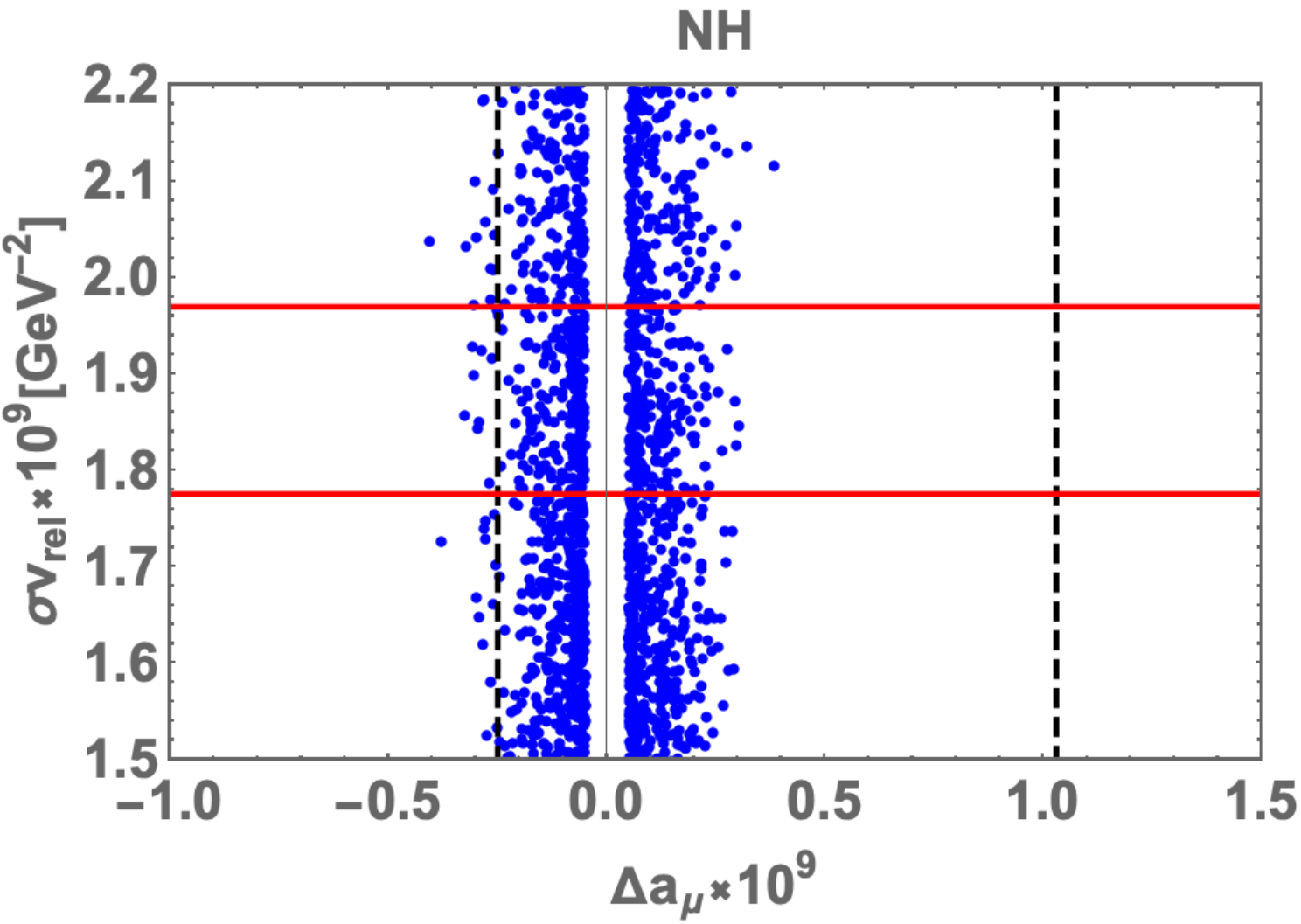}
\caption{Allowed region for the cross section for the relic density of DM and muon $g-2$ in the case of NH.
The red horizontal lines represent the upper and lower bounds corresponding to the relic density of DM in Eq.(\ref{eq:relic_exp}).
The vertical dotted lines are the upper and lower experimental bounds on muon $g-2$ in Eq.~(\ref{eq:mg2_exp}).} 
\label{fig:nh1}
\end{center}\end{figure}
%
Fig.~\ref{fig:nh1} shows allowed region for the cross section for the relic density of DM and muon $g-2$.
Our plots are shown in the blue dots, and our model satisfies both the experimental values.
The red horizontal lines represent the upper and lower bounds corresponding to the relic density of DM in Eq.(\ref{eq:relic_exp}).
The vertical dotted lines are the upper and lower experimental bounds on muon $g-2$ in Eq.~(\ref{eq:mg2_exp}).

\subsubsection{\rm IH}

\begin{figure}[tb]\begin{center}
\includegraphics[width=88mm]{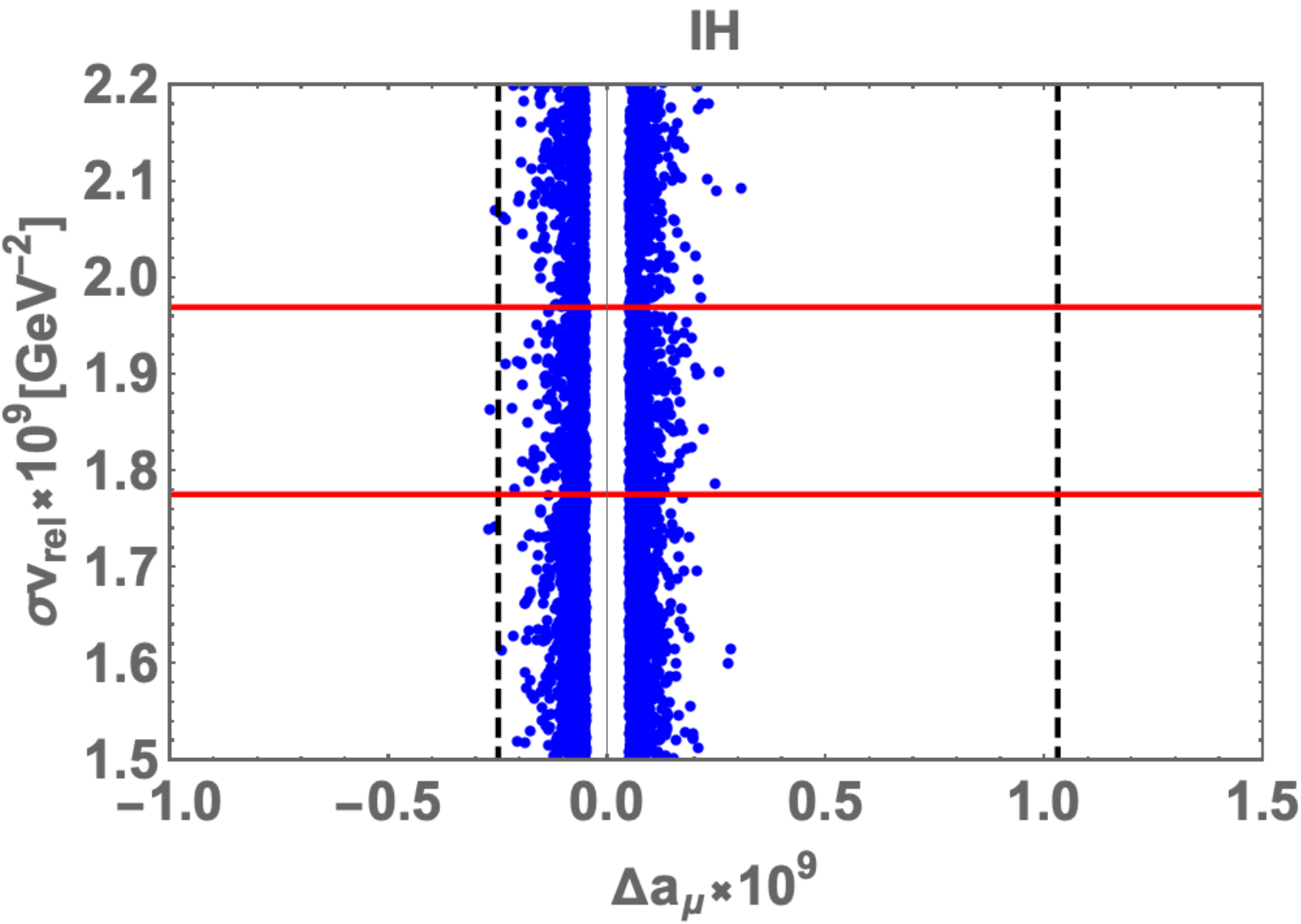}
\caption{Allowed region for the cross section for the relic density of DM and muon $g-2$
where all the legends are the same as the ones of Fig.~\ref{fig:nh1}} 
\label{fig:ih1}
\end{center}\end{figure}
%
Fig.~\ref{fig:ih1} show allowed region for the cross section for the relic density of DM and muon $g-2$ in the case of IH.
Our plots are shown in the blue dots, where all the legends are the same as the ones of Fig.~\ref{fig:nh1}.
The plot tendency of IH is almost same as the one of NH.

\section{Summary and discussion}
\label{sec:IV}

In this work, we have proposed a new mechanism in which the masses of the first and second generations in the charged-lepton sector are radiatively generated at the one-loop level, triggered by the dynamical breaking of the symmetry associated with the Ising fusion rule. While the neutrino masses are induced at the one-loop level, but the IFR remains unbroken and plays a crucial role in stabilizing the particles inside the loop. Therefore, the IFR in the neutrino sector effectively behaves as an exact 
$Z_2$ symmetry, persisting unbroken at all loop orders.

In addition to the mass generation mechanism, we have investigated several phenomenological implications of the model. Specifically, we have analyzed lepton flavor-violating processes, the anomalous magnetic dipole moment of the muon, and the relic density of dark matter. In our setup, the DM candidate is identified as a singlet scalar boson. To estimate its thermal relic density, we have computed the DM annihilation cross section by performing a velocity expansion of the annihilation amplitude in the non-relativistic regime.

Finally, we have presented the numerical results for both the DM annihilation cross section and the muon $g-2$, considering both normal hierarchy and inverted hierarchy scenarios for the neutrino mass spectrum. Our analysis shows that in both cases, the model can successfully accommodate the observed DM relic density. Moreover, the absolute value of the muon $g-2$ is predicted to reach up to $3 \times 10^{-10}$, which lies within the experimentally relevant range.

\section*{Acknowledgments}
This work is supported in part by the National Key Research and Development Program of China under Grant No.~2020YFC2201501 and the National Natural Science Foundation of China (NSFC) under Grant No. 12347103.


\bibliography{ctma4.bib}
\end{document}